\pdfoutput=1
\documentclass[a4paper,twoside]{article}
\usepackage{url}
\usepackage{hyperref}

\usepackage{epsfig}
\usepackage{graphicx}
\usepackage{subcaption}
\usepackage{calc}
\usepackage{amssymb}
\usepackage{amstext}
\usepackage{amsmath}
\usepackage{amsthm}
\usepackage{multicol}
\usepackage{pslatex}
\usepackage{booktabs}
\usepackage{apalike}
\usepackage{enumitem}
\usepackage{subcaption}
\usepackage{algorithm2e}
\usepackage[bottom]{footmisc}
\usepackage{float}
\usepackage{dblfloatfix}
\usepackage{SCITEPRESS}     

\begin{document}

\title{NavA11y: A Dynamic Analysis Approach for WCAG 2.4 Focus-Behavior Evaluation}

\author{\authorname{Abu Jafar Saifullah\orcidAuthor{0009-0005-8334-6019}, Tasmia Zerin, Zerina Begum and Kazi Sakib}
\affiliation{Institute of Information Technology, University of Dhaka, Dhaka, Bangladesh}
\email{\{bsse1109, bsse1128, zerin, sakib\}@iit.du.ac.bd}
}


\abstract{

Despite websites becoming integral to daily life, most of them are inaccessible to people with disabilities. While automated techniques exist to detect accessibility violations, prior work primarily relies on static analysis of HTML, CSS, and JavaScript. This analysis fails to determine violations which require live browser interaction. Several violations under the WCAG 2.4 focus-behavior Success Criteria (SCs), such as incorrect tab order, missing or insufficient focus indicators, and element obscuration, arise only during user interaction. To address this gap, we present a dynamic analysis approach targeting all five focus-behavior SCs defined in WCAG 2.4 navigable guidelines. The approach analyzes a webpage at two levels, page level and element level. At the page level, it drives a real browser, simulates keyboard navigation, and presses Tab to record and evaluate the focus order of elements across a webpage. At the element level, style attributes and obscuration ratio are collected for each element. Finally, these are used to check the visibility of a focus indicator and any obscuration by overlapping content. This approach is implemented as a tool called NavA11y. To evaluate this, we constructed a 22-page labeled dataset by extending six focus-related pages from the GDS Accessibility Tool Audit with 16 newly constructed test cases covering all five criteria. NavA11y detected all violations on this dataset without any false positives. To evaluate on real-world cases, we collected 26 production websites. Our approach detected 2,947 violations across those websites, with manual verification on a stratified sample confirming 90\% as true positives.

}

\onecolumn \maketitle \normalsize \setcounter{footnote}{0} \vfill

\section{Introduction}
\label{sec:introduction}

There are over one billion websites on the world wide web~\cite{internetlivestats_2024}. Yet, Web Accessibility support is largely overlooked in most websites.
An analysis conducted in 2025 revealed that of the top one million home pages, 94.8\% had accessibility errors~\cite{webaim_million_2025} with an average of 51 accessibility errors per page. 
Web Accessibility failures affect approximately 1.3 billion people (16\% of the global population) living with a disability~\cite{who_disability_2023}. This population includes those who rely on keyboard navigation as their primary means of access. A 2024 government audit of 1,151 public-sector websites found that 63.6\% failed Web Content Accessibility Guidelines (WCAG) Success Criterion~(SC)~2.1.1 (Keyboard) and 76.3\% failed SC~2.4.7 (Focus Visible)~\cite{gds_monitoring_2024}. Although many automated methods exist to detect accessibility violations, none of them cover violations related to keyboard focus behavior.

Keyboard focus behavior has two aspects, both observable only at runtime. The first is the \textit{focus state}: the visible style change a browser applies to an element on receiving focus. It is typically an outline, border change, or background highlight. A static tool can detect whether a focus state is declared in the source. However, it cannot determine whether the browser renders it during navigation, or whether the focused element is obscured by a sticky header (a fixed navigation bar) or an overlay. The second aspect is navigational: the order in which elements receive focus depending on the live Document Object Model (DOM) and active JavaScript event handlers. This order can differ substantially from what the HTML source alone implies. Both aspects leave no trace in the HTML, CSS, or Accessible Rich Internet Applications (ARIA) source. Focus behavior properties are therefore ``runtime properties that no screenshot can capture''~\cite{gena11y_he_fse2025}.

WCAG~2.4 (Navigable) defines five success criteria for keyboard focus behavior~\cite{wcag22_guideline24}: SC~2.4.3 (Focus Order), SC~2.4.7 (Focus Visible), SC~2.4.11 (Focus Not Obscured, Minimum), SC~2.4.12 (Focus Not Obscured, Enhanced), and SC~2.4.13 (Focus Appearance). Evaluating any of them requires a dynamic analysis approach that can drive a real browser, simulate keystrokes, and observe the rendered page state at each focus event. Afterwards, it should be able to accurately analyze the above-mentioned SCs.

In current literature, static tools such as WAVE~\cite{wave}, axe-core~\cite{axecore}, Lighthouse~\cite{lighthouse}, pa11y~\cite{pa11y}, and QualWeb~\cite{qualweb} inspect HTML, CSS, and ARIA attributes to detect accessibility violations. Large language model~(LLM) based tools such as GenA11y~\cite{gena11y_he_fse2025} go beyond static source inspection by reasoning over rendered screenshots and HTML source. Neither approach can observe how focus moves across a page during keyboard navigation, which is precisely what focus-behavior SCs require. BAGEL~\cite{chiou2023bagel} takes a fundamentally different approach. It drives a real browser, simulates Tab key navigation, and analyzes the rendered page layout to detect navigation barriers for SC~2.4.3 and SC~2.4.7. Its SC~2.4.3 analysis centers on whether elements are reachable via keyboard. It does not detect misuse of \texttt{tabindex} (an HTML attribute that controls focus order), mismatches between the tab order and DOM order, or cases where CSS moves elements to a different visual position than their DOM order. It also does not address focus-indicator size or contrast~(SC~2.4.13), or whether a focused element is obscured by overlapping content~(SC~2.4.11, SC~2.4.12).

This paper presents a novel dynamic analysis approach for evaluating all five focus-behavior SCs in WCAG~2.4. This approach is implemented as a tool called \textbf{NavA11y}, which drives a real browser via Playwright~\cite{playwright} to simulate keyboard navigation and observe the rendered page state at each focus event. The approach consists of two analysis strategies. Page-level analysis~(PLA) simulates Tab key traversal across the full page to evaluate focus order~(SC~2.4.3). Then it tests this traversal based on DOM order sequence, attribute misuse, and incomplete traversal. On the other hand, element-level analysis~(ELA) evaluates visibility and obscuration violations. This phase individually focuses each element and captures its rendered focus state and any obscuration by overlapping content~(SC~2.4.7, SC~2.4.11, SC~2.4.12, and SC~2.4.13). For evaluating visibility, the captured state is compared with the unfocused state to identify significant change. For obscuration, a $7 \times 7$ grid probe is used to detect any overlapping content on the focused element.

We evaluated NavA11y against a labeled dataset ($D_d$), consisting of 22 test pages with known ground-truth verdicts, and against a real-world dataset~($D_r$) of 26 production websites drawn from the Semrush top-site rankings~\cite{semrush2024}. $D_d$ originally contained 6 focus-related test pages that contained no pass cases, no boundary cases, and no obscuration cases. Hence, we extended the original $D_d$ dataset with 16 newly constructed pages to cover all five focus-behavior SCs.
NavA11y achieved 100\% precision and 100\% recall across these test pages.
Against the 26 production websites, it detected 2,947 violations across all focus-behavior SCs.
A manual evaluation of a stratified sample from six production websites was conducted to validate these findings. Two experienced front-end engineers were asked to verify the accessibility issues detected by NavA11y on these websites. Their independent assessments confirm an average of 90\% of reported violations as true positives. Therefore, these results demonstrate that our approach is a viable and effective approach to minimizing the gap in automated accessibility testing.

The paper makes the following two contributions:
\begin{itemize}
  \item A \textbf{dynamic runtime analysis approach} for detecting focus behavior violations, the first approach to cover all five WCAG~2.4 focus-behavior success criteria. We implement this approach as \textbf{NavA11y}.
  \item A \textbf{reusable focus behavior dataset} ($D_d$) containing 22 labeled test pages covering all five focus-behavior success criteria, released for evaluating future tools.
\end{itemize}

\section{Background}
\label{sec:background}
The Web Content Accessibility Guidelines~(WCAG), published by the World Wide Web Consortium~(W3C), define the international standard for web accessibility~\cite{wcag22_guideline24}. WCAG~2.2, the current version, organizes requirements into four principles. These are Perceivable, Operable, Understandable, and Robust, each subdivided into guidelines. Each guideline contains testable statements, called success criteria~(SCs), that specify measurable requirements a web page must meet. 
The criteria span three conformance levels: Level~A (the minimum baseline), Level~AA (required by most legal frameworks), and Level~AAA (the highest standard)~\cite{wcag22_guideline24}. This work focuses exclusively on the Success Criteria related to keyboard focus behavior within Guideline 2.4 (Navigable) of the WCAG 2.2 standard.

\subsection{WCAG 2.4 Navigable Guideline}
WCAG~2.2 Guideline~2.4 \textit{Navigable} contains thirteen SCs that specify how users must be able to navigate to and through page content using a keyboard~\cite{wcag22_guideline24}. These 13 criteria fall into two groups: eight concern static properties assessable from the page's static structure, and five concern keyboard focus behavior observable only at runtime~\cite{gena11y_he_fse2025}.

The static criteria cover page titles, headings, and link labels. The dynamic criteria cover keyboard focus behavior~\cite{gena11y_he_fse2025}.

\subsection{Keyboard Navigation and Focus Indicators}
\label{sec:keyboard-nav}

\begin{figure}[h!]
    \centering
    \includegraphics[width=0.65\columnwidth]{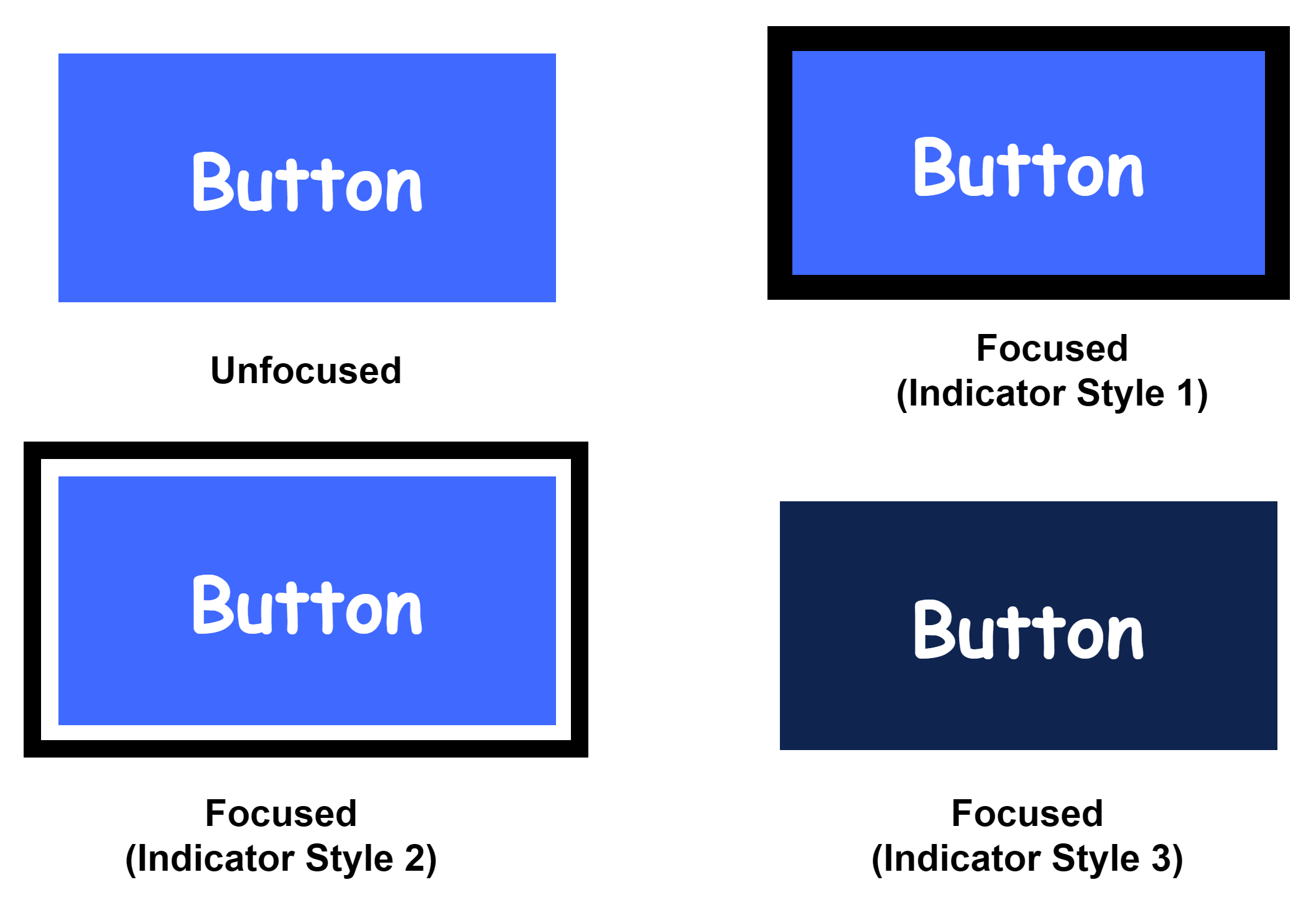}
    \caption{Example of a focused element with different indicator styles}
    \label{fig:focus-example}
\end{figure}

Keyboard navigation in browsers is driven by the Tab key, which advances focus to the next interactive element, and Shift+Tab, which moves focus in reverse~\cite{wcag22_guideline24}. The browser tracks the currently focused element and applies a visual focus indicator to it. This indicator is typically an outline, border change, or background highlight, rendered by the browser's default styles or the page's \texttt{:focus} CSS rule. 
The examples can be seen in Figure~\ref{fig:focus-example}.

The order in which elements receive focus is determined by the DOM order and any \texttt{tabindex} attributes set by the page author~\cite{wcag_technique_c27}. A \texttt{tabindex} attribute controls whether an element can receive focus and its position in the Tab sequence. Both the rendered indicator and the focus sequence exist only in the live browser state and are absent from the HTML source.

\subsection{Focus-Behavior Success Criteria}
\label{sec:focus-scs}

The five focus-behavior SCs listed in Table~\ref{tab:focus-scs} test properties that emerge only after a browser renders the page and responds to keyboard input. The following paragraphs describe each criterion.\\

\begin{table}[ht]
  \centering
  \small
  \setlength{\tabcolsep}{4pt}
  \caption{Five WCAG~2.4 focus-behavior Success Criteria evaluated by NavA11y.}
  \label{tab:focus-scs}
  \renewcommand{\arraystretch}{1.2}
  \begin{tabular}{|p{0.95cm}|p{4.2cm}|c|}
    \hline
    \textbf{SC} & \textbf{Full Name} & \textbf{Level} \\
    \hline
    2.4.3  & Focus Order                   & A   \\ \hline
    2.4.7  & Focus Visible                 & AA  \\ \hline
    2.4.11 & Focus Not Obscured (Minimum)  & AA  \\ \hline
    2.4.12 & Focus Not Obscured (Enhanced) & AAA \\ \hline
    2.4.13 & Focus Appearance              & AAA \\ \hline
  \end{tabular}
  \vspace{-1em}
\end{table}

\textbf{SC~2.4.3 Focus Order.}
This criterion requires that a web page can be navigated sequentially using the keyboard and the interactive elements should receive focus in an order that preserves meaning and operability~\cite{wcag22_guideline24}. Browsers determine this order from the sequence in which elements appear in the HTML source (DOM order), but CSS positioning can cause the visual layout to diverge from the source order, creating a mismatch between what the user sees and where keyboard focus moves~\cite{wcag_technique_c27}.

\textbf{SC~2.4.7 Focus Visible.}
This criterion requires that when an interactive element receives keyboard focus, a visible focus indicator is displayed~\cite{wcag22_guideline24}. The indicator is typically an outline, border change, or background color shift rendered by the browser's default styles or the page's CSS. Indicator appearance depends on how the browser styles the element after focus is triggered. This cannot be determined from the HTML source.

\textbf{SC~2.4.11 Focus Not Obscured, Minimum (Level~AA).}
This criterion requires that when an element receives keyboard focus, it is not entirely hidden by author-created content such as sticky headers, fixed footers, or cookie banners~\cite{wcag22_guideline24}. At minimum, some part of the focused element must remain visible so the user can locate their position. This can only be determined at runtime.

\textbf{SC~2.4.12 Focus Not Obscured, Enhanced (Level~AAA).}
This criterion strengthens SC~2.4.11 by requiring that no part of a focused element is hidden by other content~\cite{wcag22_guideline24}. While SC~2.4.11 permits partial obstruction as long as some portion remains visible, SC~2.4.12 requires an element to be fully unobscured. The same runtime constraints apply as in SC~2.4.11.

\textbf{SC~2.4.13 Focus Appearance (Level~AAA).}
This criterion requires that the focus indicator meets minimum size and contrast requirements, i.e., the indicator must be at least 2~CSS pixels thick and have a contrast ratio of at least 3:1 against the unfocused state and against adjacent colors~\cite{wcag22_guideline24}. Measuring the indicator's size and color can only be done after the browser has rendered the page and applied focus.


\section{Approach}
\label{sec:approach}

The goal of our approach is to evaluate the five focus-behavior SCs listed in Table~\ref{tab:focus-scs}.

\begin{figure}[ht]
    \centering
    \includegraphics[width=0.95\columnwidth]{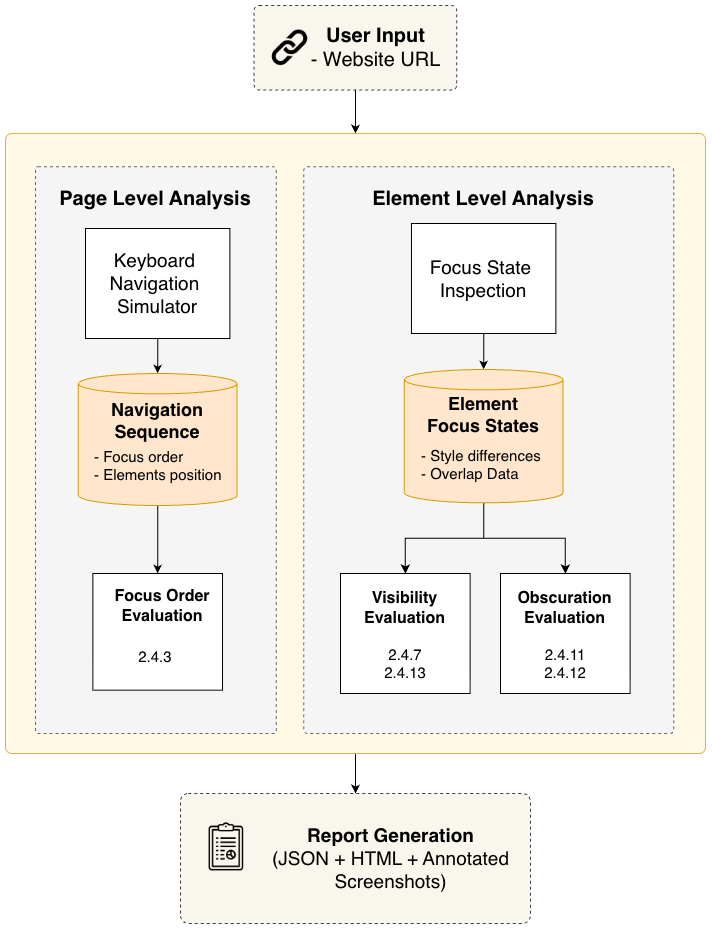}
    \caption{Overview of the NavA11y dynamic analysis architecture.}
    \label{fig:Architecture}
\end{figure}

We implement it as a tool named \textbf{NavA11y}. Given a webpage URL, a browser session is launched and two separate analyses are conducted: Page-Level Analysis and Element-Level Analysis.
Page-Level Analysis evaluates SC~2.4.3 (Focus Order). Element-Level Analysis evaluates SC~2.4.7, SC~2.4.11, SC~2.4.12, and SC~2.4.13. The two phases are structurally distinct. SC~2.4.3 requires observing the order in which focus traverses the entire page, which demands a full keyboard traversal. The remaining four SCs require observing each element's rendered focus state and obscuration individually, which demands direct per-element inspection.\\
For Page-Level Analysis, Tab key presses are simulated across the page to record the focus path. For Element-Level Analysis, each focusable element, i.e., any element that can receive keyboard focus (links, buttons, and form inputs), is targeted directly to capture its rendered state. The outputs from these phases are passed to the corresponding evaluation checks, as shown in Figure~\ref{fig:Architecture}. After evaluation, accessibility reports are generated for the webpage. The following subsections describe each phase in detail.

\subsection{Page-Level Analysis}
\label{sec:focus-order}

As defined in Section~\ref{sec:background}, SC~2.4.3 requires that focus moves through the page in an order that preserves meaning and operability~\cite{wcag22_guideline24}. When this order is disrupted, keyboard-only users lose track of their position on the page~\cite{wcag22_understanding_243}. Such disruptions are detected by recording three parallel orderings of a page's focusable elements and testing for divergence between them.

\subsubsection{Keyboard Navigation Simulator}
A \textit{Navigation Sequence} is recorded, composed of three parallel orderings: DOM order, tab sequence, and visual reading order. The three orderings together cover all known SC~2.4.3 violation patterns~\cite{wcag22_guideline24,wcag_technique_c27}. The tab sequence may not follow the HTML source, the visual layout may not match either the HTML source or the tab sequence, or all three may disagree simultaneously.
\textit{DOM order} lists every focusable element in the sequence as it appears in the HTML source. It is collected before any Tab key is pressed and serves as the baseline against which the other two orderings are compared.\\
\textit{Tab sequence} records the order in which elements receive focus during a simulated keyboard traversal. The traversal stops when focus returns to the start, when a focus trap is detected, or when a limit of 200~key presses is reached. This limit was validated empirically: a pilot study across $D_r$ (the real-world dataset defined in Section~\ref{sec:eval-setup}) measured a 95th-percentile DOM focusable element count of 178~elements. The cap of 200 exceeds P95, so pages up to the 95th percentile are covered in full. Only the largest outliers may be truncated. Three consecutive repeats on the same element are flagged as a focus trap. One or two repeats can occur naturally when navigation wraps or a page has few elements, but three consecutive repeats with no forward progress indicate a deadlock.\\
\textit{Visual reading order} ranks elements by their screen position, left to right and top to bottom~\cite{wcag_technique_c27}. It captures the spatial layout the user sees, independent of source order or keyboard behavior. These three orderings form the Navigation Sequence and are evaluated next.

\subsubsection{Focus Order Evaluation}
To evaluate focus order, our approach tests the Navigation Sequence against five independent checks. Multiple checks are needed because SC~2.4.3 violations arise from structurally different causes. Four checks each target a distinct violation pattern: rank divergence, attribute misuse, CSS repositioning, and incomplete traversal. A fifth check guards against issuing a verdict when the statistical basis is too small to be reliable. Their conditions and thresholds are shown in Table~\ref{tab:dynamic-thresholds}.

\textit{\textbf{Order divergence.}} DOM order is compared with the tab sequence using Kendall~$\tau$~\cite{kendall1938}, a statistical measure of how similar two ranked lists are: 1.0 means identical order; lower values indicate increasing divergence.\\
To detect significant deviations in focus order, we define a failure threshold of $\tau < 0.90$, which identifies substantial disruptions to the logical navigation sequence. This threshold tolerates minor focus-order variations that can arise at runtime, such as those caused by JavaScript inserting or rearranging a small number of elements after page load~\cite{wcag22_understanding_243}. To validate this choice, we conducted a pilot study across all 26~sites in $D_r$ (described in Section~\ref{sec:evaluation}). All 26 sites completed traversal; 23 had perfect agreement ($\tau = 1.0$). The three sites with non-zero divergence ebay ($\tau = 0.989$), yahoo ($\tau = 0.751$), and qualtrics ($\tau = 0.685$) are each flagged as SC~2.4.3 violations in the production evaluation (Section~\ref{sec:results-dr}). Naturally occurring $\tau$ therefore stays at or above 0.989, giving the 0.90~threshold a margin of approximately 0.089 above the observed baseline.\\
\textit{\textbf{Positive \texttt{tabindex}.}} Any element with a positive \texttt{tabindex} is flagged~\cite{wcag_technique_f44}, as it overrides DOM order and disrupts the expected navigation sequence. One exception is skip links~\cite{wcag_technique_g1}. Anchors whose \texttt{href} begins with \texttt{\#} and whose \texttt{id} or \texttt{class} matches a known pattern (e.g., \texttt{skip}, \texttt{skip-to-main}) are excluded, since WCAG explicitly recommends their use~\cite{wcag_technique_g1}.

\textit{\textbf{Truncated sequence.}} If a Tab traversal covers fewer than 60\% of a page's focusable elements, the result is flagged as truncated. This threshold is set by reasoned argument: at 60\% coverage, only approximately 36\% of all pairwise element relationships are captured (pair counts scale quadratically with element count), leaving the Kendall~$\tau$ score statistically unreliable.

\textit{\textbf{Small-page guard.}} Pages with five or fewer focusable elements are routed to manual review (Table~\ref{tab:dynamic-thresholds}). The threshold of five is set by reasoned argument: at N\,=\,5, a single transposed element shifts the Kendall~$\tau$ score by 0.2~\cite{kendall1938}, making every plausible ranking statistically indistinguishable from a violation. For both the truncated-sequence and small-page conditions, no pass or fail verdict is issued. Instead, a warning is recorded in the report to flag the page for manual review.

\textit{\textbf{Visual-order mismatch.}} CSS can visually move an element to a different position on screen without changing where it sits in the HTML source. When this happens, the Tab key follows the HTML order, but the user sees elements in a different order. This is detected by computing Kendall~$\tau$ between the visual reading order and the tab sequence. A score below the same divergence threshold triggers the flag because a keyboard user cannot predict where focus will land next from what they see on screen~\cite{wcag_technique_c27}.

\begin{table}[t]
  \centering
  \small
  \caption{Evaluation conditions with their thresholds.}
  \label{tab:dynamic-thresholds}
  \begin{tabular}{@{}llll@{}}
    \toprule
    \textbf{SC} & \textbf{Condition} & \textbf{Threshold} & \\
    \midrule
    2.4.3  & Kendall $\tau$ divergence  & $\tau < 0.90$           \\
           & Positive \texttt{tabindex} & Any value $> 0$          \\
           & Truncated sequence         & $< 60\%$ covered         \\
           & Small-page guard           & $\leq 5$ elements        \\
    \midrule
    2.4.7  & Focus indicator            & Any visual change        \\
    \midrule
    2.4.11 & Obscuration                & $\geq 99\%$ covered     \\
    2.4.12 & Obscuration                & Any coverage             \\
    \midrule
    2.4.13 & Indicator thickness        & $\geq 2$\,CSS\,px      \\
           & Indicator contrast         & $\geq 3\!:\!1$ ratio   \\
    \bottomrule
  \end{tabular}
\end{table}

\subsection{Element-Level Analysis}
\label{sec:focus-visibility}

The four element-level SCs check two properties of each focusable element. First, whether a visible focus indicator appears. Second, whether the element is obscured by any overlapping content such as a sticky header or cookie banner.
Keyboard users depend on these properties to identify their position on the page~\cite{wcag22_guideline24}. As established in Section~\ref{sec:background}, both properties are absent from the HTML source and can only be observed at runtime~\cite{wcag_technique_f78,wcag22_guideline24}.
\subsubsection{Focus State Inspection}
For each focusable element, a pre-focus snapshot of computed CSS properties is captured using \texttt{getComputedStyle()}~\cite{mdn_getcomputedstyle}. The element is then focused using the \texttt{page.focus()} API from Playwright~\cite{playwright}, a browser-automation library, and a post-focus snapshot is captured after a 300\,ms delay. Property selection was driven by three WCAG source documents: Technique~F78 (a documented failure pattern for CSS-hidden focus indicators)~\cite{wcag_technique_f78}, and the understanding documents for SC~2.4.7~\cite{wcag22_guideline24} and SC~2.4.13~\cite{wcag22_guideline24}. Every CSS property named in those documents in the context of focus indicator appearance or removal was extracted. This yielded six property groups spanning 22 properties: \textit{outline} (style, width, color, offset), \textit{box-shadow}, \textit{border} (per-side width and color), \textit{background and text color}, \textit{text decoration} (line, style, color, thickness), and \textit{transform, opacity, filter, visibility, and display}.

The 300\,ms delay provides a conservative margin for CSS transitions to settle before the post-focus snapshot is captured~\cite{css_transitions_level1}. A $7 \times 7$ grid of 49 evenly spaced points is then placed over the element's screen area. At each point, the browser's \texttt{elementsFromPoint()} API~\cite{mdn_elementsfrompoint} returns the list of elements at that location, ordered from top to bottom. The topmost element at each point is checked to see whether it is the focused element or overlapping content.

The $7 \times 7$ grid size was selected through a pilot study on $D_r$ that tested four grid densities against a $10 \times 10$ pixel-level reference across 412 on-screen elements. The study started at $5 \times 5$ and increased density until the first grid that reached $\geq 99\%$ agreement on both SC verdicts. As Table~\ref{tab:grid-density} shows, $6 \times 6$ fell short for SC~2.4.12 at 98.30\%. The $7 \times 7$ grid was the first to cross that threshold at 99.03\%, using 51\% fewer probe points than the $10 \times 10$ reference (49 vs.\ 100).

\begin{table}[t]
  \centering
  \small
  \caption{Grid density comparison across 412 on-screen elements from $D_r$, measured against the $10 \times 10$ reference.}
  \label{tab:grid-density}
  \begin{tabular}{lccc}
    \toprule
    \textbf{Grid} & \textbf{Probes} & \textbf{SC~2.4.11 agree} & \textbf{SC~2.4.12 agree} \\
    \midrule
    $5 \times 5$   &  25 & 100.00\% & 94.42\% \\
    $6 \times 6$   &  36 & 100.00\% & 98.30\% \\
    $7 \times 7$   &  49 & 100.00\% & 99.03\% \\
    $8 \times 8$   &  64 & 100.00\% & 99.51\% \\
    $10 \times 10$ & 100 & \multicolumn{2}{c}{(reference)} \\
    \bottomrule
  \end{tabular}
  \vspace{-0.5em}
\end{table}

\begin{figure}[ht]
    \centering
    \includegraphics[width=0.95\columnwidth]{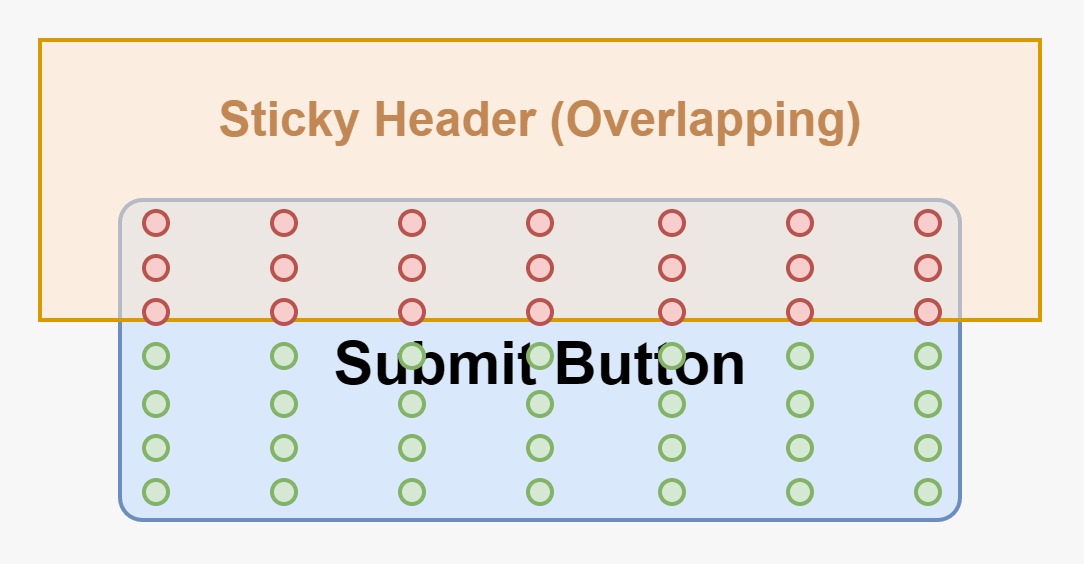}
    \caption{The $7 \times 7$ grid probe for obscuration detection. Green points are unobscured; red points are covered by a sticky header.}
    \label{fig:obscuration-probe}
\end{figure}

The collected style snapshots and obscuration data are evaluated by two independent checks.

\subsubsection{Visibility Evaluation}
The Visibility Evaluation determines whether a detectable focus indicator appeared when the element received focus. 
Without any focus indicator, a keyboard user cannot identify which element is currently active~\cite{wcag22_guideline24}.
Pre-focus and post-focus style snapshots are compared across the six property groups identified in Section~\ref{sec:focus-visibility}: outline, box-shadow, border, background and text color, text decoration, and transform and opacity. We introduce a three-tier confidence classification to distinguish the strength of evidence for each detected change:
\begin{itemize}
  \item \textit{High}: a change to a dedicated focus-indicator property identified by WCAG Technique~F78~\cite{wcag_technique_f78}, such as \texttt{outline} or \texttt{border} (e.g., a new outline appears on focus).
  \item \textit{Medium}: a change to a secondary visual property that can serve as a perceivable indicator but is not a dedicated focus style, such as background color or text decoration~\cite{wcag22_guideline24}.
  \item \textit{Low}: a change to a property not visually associated with focus indication, such as a CSS transform translation with no accompanying color change.

\end{itemize}
SC~2.4.7 (Level~AA) passes if any medium or high-confidence change is found, since the criterion only requires that a focus indicator ``is visible''~\cite{wcag22_guideline24}. It fails when only low-confidence changes or no changes are detected. It also fails if a style change removes an existing indicator without adding a new one. This is a failure pattern documented in WCAG Technique~F78~\cite{wcag_technique_f78}. SC~2.4.13 (Level~AAA) applies stricter rules from the criterion~\cite{wcag22_guideline24}. The indicator must be at least 2~CSS pixels thick and achieve a contrast ratio of at least 3:1 against the unfocused state (Table~\ref{tab:dynamic-thresholds}). The second element-level check evaluates whether overlapping page content obscures the focused element from view, regardless of whether an indicator is present.

\begin{figure*}[ht]
    \centering
    \begin{subfigure}[b]{1\columnwidth}
        \centering
        \includegraphics[width=1.1\columnwidth]{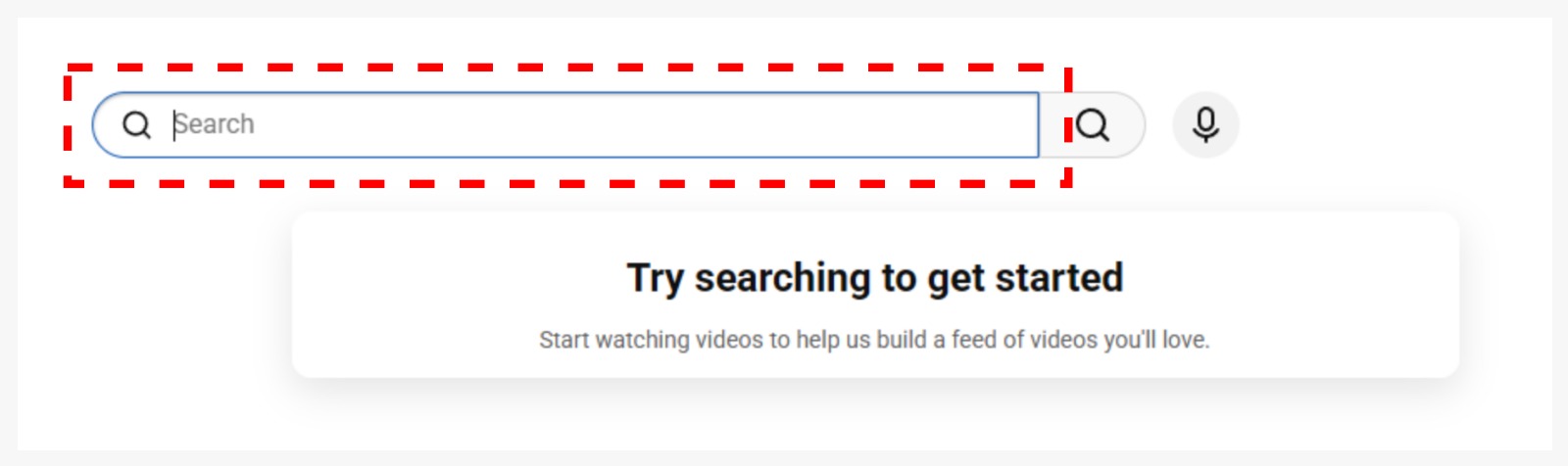}
        \caption{Violation when focused on search bar (SC~2.4.7)}
        \label{fig:focus-visible-UI}
    \end{subfigure}\hfill
    \begin{subfigure}[b]{1\columnwidth}
        \centering
        \includegraphics[width=0.9\columnwidth]{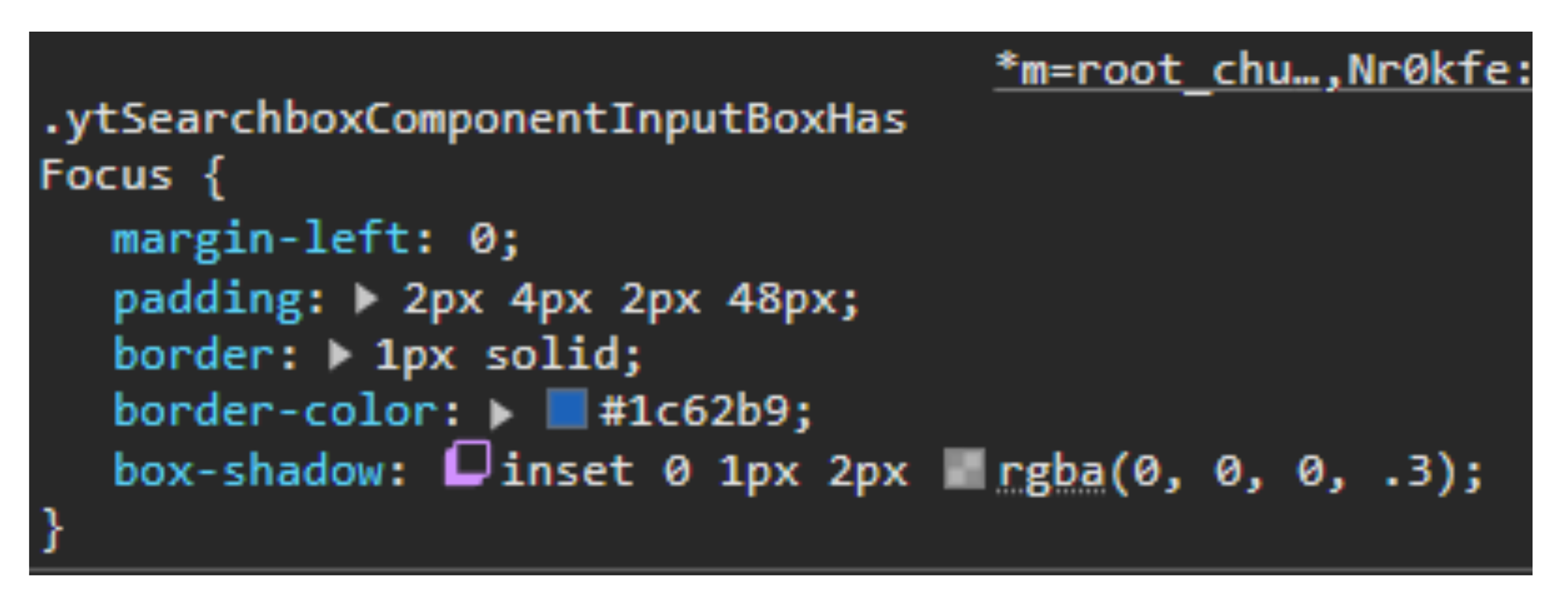}
        \caption{CSS styles responsible for violation (SC~2.4.13)}
        \label{fig:focus-code}
    \end{subfigure}
    \vspace{.5em}
    \caption{SC~2.4.7 and SC~2.4.13 violations on youtube.com: focused search input exhibits barely visible change in outline (SC~2.4.7), therefore also fails the stricter 2\,px thickness and 3:1 contrast requirement of SC~2.4.13.}
    \label{fig:focus-visible-side-by-side}
\end{figure*}

\subsubsection{Obscuration Evaluation}
The Obscuration Evaluation determines whether overlapping page content hides the focused element from the user.
Sticky headers, fixed footers, and cookie consent banners commonly overlay page content as the user scrolls, and can cover a focused element so that the user cannot see it or its focus indicator~\cite{wcag22_guideline24}.
The $7 \times 7$ grid probe is used to measure such obscuration. Each of the 49 points reveals whether the topmost element there belongs to the focused element or to overlapping content. The share of obscured points is the element's obscuration ratio. SC~2.4.11 (Level~AA) fails only when this ratio reaches 99\%, which approximates the ``entirely hidden'' condition in the criterion~\cite{wcag22_guideline24} while leaving a small margin for pixel rounding. SC~2.4.12 (Level~AAA) fails if even one point is obscured, since the criterion requires the element to be fully visible~\cite{wcag22_guideline24}.

\begin{figure}[ht]
    \centering
    \includegraphics[width=0.8\columnwidth]{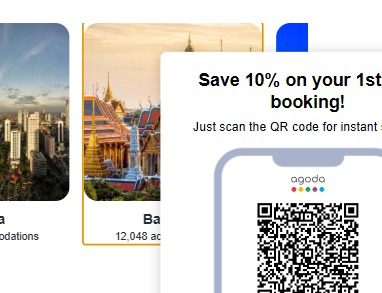}
    \caption{SC~2.4.11 and SC~2.4.12 violations on agoda.com: a focused article card partially hidden beneath a sticky header. SC~2.4.11 fails when the element is entirely hidden; SC~2.4.12 fails as soon as any part is obscured.}
    \label{fig:obscured-fail}
\end{figure}

\subsection{Report Generation}
\label{sec:reporting}
Each finding records the page URL, SC, affected element, and supporting evidence. NavA11y produces three output formats: an HTML report, a machine-readable JSON file, and annotated screenshots.

\section{Evaluation}
\label{sec:evaluation}

This section reports an empirical evaluation of NavA11y's detection accuracy, structured around one research question.

\textbf{RQ:} How accurately does NavA11y detect focus-related WCAG~2.4 violations?

It is evaluated on a controlled dataset~($D_d$) with known verdicts and on production websites~($D_r$) for real-world performance.

\subsection{Experimental Setup}
\label{sec:eval-setup}

NavA11y\footnote{Source code available at \href{https://github.com/jaf107/NavA11y}{https://github.com/jaf107/NavA11y}.},
, which implements the approach described in Section~\ref{sec:approach}, is built in Node.js and uses Playwright~v1.58~\cite{playwright}. It runs Chromium in headless mode (browser execution without a graphical user interface), making it suitable for automated pipelines~\cite{playwright}. 
Three injected scripts capture CSS properties, focus sequence, and obscuration data per page.

\textbf{$D_d$ — WCAG~2.4 Focus-Behavior Dataset.}
The UK Government Digital Service~(GDS) provides an open-source accessibility dataset~\cite{gdsa11yaudit}. Its six focus-related pages cover only a subset of the five SCs and contain no PASS cases, boundary cases, or obscuration scenarios. These were extended with 16 author-constructed pages, giving $D_d$ 22 pages and 32 SC-level assertions spanning all five SCs. Each page was built using the sufficient techniques and failure conditions specified in WCAG~\cite{wcag22_guideline24}, following the GDS methodology~\cite{gdsa11yaudit}.

Three categories of test case are represented in the author-constructed pages. PASS cases are included for SC~2.4.3, SC~2.4.7, and SC~2.4.13 to verify that no false positives are raised. SCs~2.4.11 and~2.4.12 have no dedicated PASS pages, since any page without persistent overlays satisfies both criteria by definition~\cite{wcag22_guideline24}. Boundary cases place the page at the exact compliance threshold. For example, one boundary case uses an outline width of exactly 2\,CSS pixels, the minimum required by SC~2.4.13~\cite{wcag22_guideline24}. Obscuration cases cover fixed footers and cookie banners, exercising SCs~2.4.11 and~2.4.12. The dataset can be reused to benchmark future tools.

\textbf{$D_r$ — Real-World Validation Dataset.}
A small controlled dataset is insufficient to reveal whether NavA11y produces false positives on complex, dynamic production pages. $D_r$ assesses whether precision is maintained on sites representative of those encountered in practice. The evaluation uses the 30-site Semrush top-website corpus introduced in~\cite{gena11y_he_fse2025} and sourced from Semrush's global traffic ranking~\cite{semrush2024}. These sites represent high-traffic, professionally maintained pages spanning a range of categories (e.g., banking, e-commerce, fashion). Reusing the same corpus places NavA11y's findings in the same benchmark context as~\cite{gena11y_he_fse2025}. Four of the 30 sites were inaccessible to automated evaluation: craigslist.org (geographic IP restriction), doordash (bot-detection pop-up), makemytrip (HTTP~403 response), and sciencedirect (bot-detection blocking automated navigation), leaving 26 evaluable sites.

\paragraph{Evaluation scope.}
Each site is evaluated at its home URL using Tab key navigation only, following the same conditions as~\cite{gena11y_he_fse2025}. No login or pre-interaction is applied. Inner pages and authenticated states are not evaluated.

\subsection{Results on $D_d$}
\label{sec:results-da}
NavA11y correctly classifies all 32 SC-level assertions on $D_d$: 14 true positives~(TP), 18 true negatives~(TN), 0 false positives~(FP), and 0 false negatives~(FN). We report at the assertion level rather than the page level to reveal whether any individual SC is systematically weaker than others. Some pages assert against multiple SCs (e.g., a single obscuration page asserts against both SC~2.4.11 and SC~2.4.12). Each page carries a \texttt{data-expected} attribute. NavA11y's output was compared against this attribute to classify each assertion. Table~\ref{tab:confusion} shows the overall classification summary. Table~\ref{tab:persc} breaks down results per SC. Precision denotes the fraction of reported violations that correspond to genuine failures. Recall denotes the fraction of genuine failures that are detected; $F_1$ is their harmonic mean.

\begin{table}[H]
  \centering
  \caption{Confusion matrix for NavA11y on $D_d$ (32 assertions across 22 test pages).}
  \label{tab:confusion}
  \resizebox{\columnwidth}{!}{%
  \begin{tabular}{lcc}
    \toprule
    & \textbf{Predicted Violation} & \textbf{Predicted Pass} \\
    \midrule
    \textbf{Actual Violation} & 14 (TP) & 0 (FN) \\
    \textbf{Actual Pass}      & 0 (FP)  & 18 (TN) \\
    \bottomrule
  \end{tabular}%
  }
  \\ \smallskip
  \small Overall: Precision = 100\%, Recall = 100\%, $F_1 = 1.00$, Accuracy = 100\%.
\end{table}

As noted in Section~\ref{sec:threats}, 16 of the 22 pages are author-constructed.

\begin{table}[H]
  \centering
  \caption{Per-SC results on $D_d$.}
  \label{tab:persc}
  \resizebox{\columnwidth}{!}{%
  \begin{tabular}{lcccccc}
    \toprule
    \textbf{SC} & \textbf{TP} & \textbf{TN} & \textbf{FP} & \textbf{FN} & \textbf{Precision} & \textbf{Recall} \\
    \midrule
    2.4.3 Focus Order           & 4  & 1  & 0 & 0 & 100\% & 100\% \\
    2.4.7 Focus Visible         & 2  & 13 & 0 & 0 & 100\% & 100\% \\
    2.4.11 Not Obscured (Min)   & 2  & 0  & 0 & 0 & 100\% & 100\% \\
    2.4.12 Not Obscured (Enh)   & 2  & 0  & 0 & 0 & 100\% & 100\% \\
    2.4.13 Focus Appearance     & 4  & 4  & 0 & 0 & 100\% & 100\% \\
    \midrule
    \textbf{Total}              & \textbf{14} & \textbf{18} & \textbf{0} & \textbf{0} & \textbf{100\%} & \textbf{100\%} \\
    \bottomrule
  \end{tabular}%
  }
\end{table}

Table~\ref{tab:persc} shows that all five SCs achieve 100\% precision and 100\% recall, yielding $F_1 = 1.00$. Section~\ref{sec:results-dr} examines whether this precision extends to production websites.

\subsection{Results on $D_r$}
\label{sec:results-dr}

NavA11y detected \textbf{2,947 violations} across all 26~$D_r$ sites. Table~\ref{tab:persite} shows the per-site breakdown. Table~\ref{tab:production} summarizes aggregate totals by SC.

\begin{table}[!tb]
  \centering
  \caption{Violations detected per website in $D_r$, grouped by SC.}
  \label{tab:persite}
  \resizebox{\columnwidth}{!}{%
  \footnotesize
  \begin{tabular}{clrrrrrr}
    \toprule
    \textbf{\#} & \textbf{Website} & \textbf{2.4.3} & \textbf{2.4.7} & \textbf{2.4.11} & \textbf{2.4.12} & \textbf{2.4.13} & \textbf{Total} \\
    \midrule
     1 & adp            & 0 &  4 &  4 &  13 &   67 &   88 \\
     2 & agoda          & 1 &  4 & 26 &  62 &   41 &  134 \\
     3 & caliente       & 1 & 16 &  0 &   2 &   58 &   77 \\
     4 & capitalone     & 0 &  8 &  1 &   1 &   73 &   83 \\
     5 & cricbuzz       & 0 &  0 & 36 &  37 &  185 &  258 \\
     6 & discord        & 0 &  1 &  4 &   5 &   27 &   37 \\
     7 & doubleclick    & 0 &  0 &  0 &   5 &   13 &   18 \\
     8 & ebay           & 1 &  1 &  0 &   3 &  105 &  110 \\
     9 & fragrantica    & 0 &  3 &  1 &  91 &  \textbf{365} &  \textbf{460} \\
    10 & genius         & 1 & \textbf{39} & 32 &  32 &  141 &  245 \\
    11 & google         & 0 &  2 &  0 &   1 &   20 &   23 \\
    12 & live           & 0 &  0 &  0 &   0 &    7 &    7 \\
    13 & nih.gov        & 0 &  0 &  0 &   0 &    3 &    3 \\
    14 & openai         & 0 &  1 &  0 &   2 &   76 &   79 \\
    15 & progressive    & 0 &  1 &  0 &   0 &    2 &    3 \\
    16 & qualtrics      & 1 & 18 &  6 &  28 &  112 &  165 \\
    17 & samsung        & 0 &  0 &  0 &   2 &   19 &   21 \\
    18 & shein          & 1 &  5 & \textbf{78} &  \textbf{80} &    5 &  169 \\
    19 & stackoverflow  & 0 &  0 & 43 &  73 &  114 &  230 \\
    20 & steamcommunity & 0 &  3 &  8 &  11 &  190 &  212 \\
    21 & usps           & 0 & 15 &  3 &   4 &   33 &   55 \\
    22 & walmart        & 0 &  2 & 57 &  57 &   54 &  170 \\
    23 & wikipedia      & 0 &  0 &  0 &   0 &   30 &   30 \\
    24 & yahoo          & 1 & 25 & 27 &  33 &   79 &  165 \\
    25 & youtube        & 0 &  8 &  2 &   7 &   12 &   29 \\
    26 & zerodha        & 0 &  0 &  0 &   0 &   76 &   76 \\
    \midrule
       & \textbf{Total}     & \textbf{7} & \textbf{156} & \textbf{328} & \textbf{549} & \textbf{1,907} & \textbf{2,947} \\
    \bottomrule
  \end{tabular}
  }
\end{table}

\begin{table}[H]
  \centering
  \caption{Aggregate violations on $D_r$ (26 production websites) by SC.}
  \label{tab:production}
  \resizebox{\columnwidth}{!}{%
  \begin{tabular}{llcc}
    \toprule
    \textbf{SC} & \textbf{Name} & \textbf{Violations} & \textbf{Sites Affected} \\
    \midrule
    2.4.3  & Focus Order           &      7 &  7/26 (27\%) \\
    2.4.7  & Focus Visible         &    156 & 18/26 (69\%) \\
    2.4.11 & Not Obscured (Min)    &    328 & 15/26 (58\%) \\
    2.4.12 & Not Obscured (Enh)    &    549 & 21/26 (81\%) \\
    2.4.13 & Focus Appearance      & \textbf{1,907} & \textbf{26/26 (100\%)} \\
    \midrule
    \textbf{Total} &              & \textbf{2,947} & \textbf{26/26 (100\%)} \\
    \bottomrule
  \end{tabular}%
  }
\end{table}

SC~2.4.13 (Focus Appearance) accounts for 65\% of all violations (1,907 of 2,947), affecting all 26 evaluated sites. Many production websites rely on the browser-default \texttt{:focus} outline without customizing it~\cite{webaim_million_2024}. These defaults typically render as a thin 1--2\,px outline that commonly falls below the required 2\,CSS pixel thickness or 3:1 contrast ratio~\cite{wcag22_guideline24}. SC~2.4.12 (Focus Not Obscured Enhanced) is the second most prevalent, accounting for 19\% of violations (549 across 21 sites), consistent with the persistent overlay patterns described in Section~\ref{sec:approach}.

Violation counts range from 3 (nih.gov) to 460 (fragrantica). Sites with elevated SC~2.4.11 and SC~2.4.12 counts are shein (SC~2.4.11: 78, SC~2.4.12: 80) and walmart (SC~2.4.11: 57, SC~2.4.12: 57). Both websites employed persistent overlays that obstructed numerous focusable elements during the evaluation run. Sites with elevated SC~2.4.13 counts are fragrantica (365) and cricbuzz (185). Both of these websites depended on browser-default focus styles that lacked sufficient indicator contrast or thickness during the evaluation run.

Three categories of edge case affect the $D_r$ results. First, on some sites, JavaScript event handlers trapped or redirected focus before all focusable elements were reached. NavA11y flags these truncated traversals with an advisory notice rather than a pass or fail verdict.
Second, dynamic overlays such as coupon dialogs and newsletter pop-ups can appear during Tab traversal and cover focused elements. Because these overlays may appear only on first load or after a timed delay, violation counts for affected sites can vary across runs. The 2,947 violations in Table~\ref{tab:persite} reflect a single benchmark run.
Third, infinite-scroll pages mutate the DOM during traversal, which corrupts style comparisons between pre-focus and post-focus snapshots. This structural limitation is discussed in Section~\ref{sec:threats}.

To assess precision, six sites were selected from $D_r$ using tercile stratification (two per tercile). These were: yahoo and agoda (high), walmart and usps (mid), and zerodha and nih.gov (low). Each reported violation was manually verified against the corresponding WCAG SC definition. For each violation, the raters received the element selector (a CSS path uniquely identifying the element in the DOM), a screenshot of the focused element, and the WCAG SC identifier. They judged whether the reported condition constituted a genuine violation of that criterion.

Prior to labeling, 62 reported violations on yahoo spanning all four element-level SCs (SC~2.4.7, SC~2.4.11, SC~2.4.12, and SC~2.4.13) were identified as infinite-scroll artifacts and excluded. The exclusion criterion was applied systematically. A reported violation was classified as an infinite-scroll artifact if and only if the obscuring element or changed style property was absent from the pre-traversal DOM snapshot captured at page load, indicating it was introduced by a scroll-triggered content load. The remaining 541 reported violations constitute the labeling dataset.

Two experienced front-end engineers independently validated the detected accessibility issues and labeled them as true positives or false positives. The evaluators agreed on 487 out of 541 cases, resulting in an observed agreement of 90\%. Cohen’s $\kappa$ was calculated to assess inter-rater reliability. However, due to the highly imbalanced distribution of labels (i.e., most issues were classified as true positives), $\kappa$ was unstable and yielded a negative value. Therefore, we report the observed agreement and confusion matrix to better reflect the consistency between evaluators.

\begin{table}[h]
\centering
\caption{Inter-rater agreement matrix for manual precision labeling ($N=541$).}
\label{tab:cohen-kappa}
\begin{tabular}{lcc}
\hline
\textbf{Rater 1 / Rater 2} & \textbf{True Positive} & \textbf{False Positive} \\
\hline
True Positive  & 486 & 33 \\
False Positive &  21 &  1 \\
\hline
\end{tabular}
\smallskip
\noindent $P_o = 0.9002$,\quad $P_e = 0.9016$,\quad $\kappa = -0.0142$
\end{table}

The negative $\kappa$ is a known artifact of extreme class imbalance: when one category dominates, $P_e$ approaches $P_o$ and $\kappa$ becomes unstable~\cite{byrt1993bias}. To obtain a prevalence-adjusted measure, we compute the Prevalence-Adjusted Bias-Adjusted Kappa (PABAK)~\cite{byrt1993bias}:
\[
  \text{PABAK} = 2P_o - 1 = 2 \times 0.9002 - 1 = 0.80
\]
A PABAK of 0.80 indicates substantial agreement on the Landis and Koch scale~\cite{landis1977measurement}, confirming that the two raters reached consistent verdicts despite the skewed label distribution.

\section{Related Work}
\label{sec:related}

To our knowledge, no existing tool covers all five WCAG~2.4 focus-behavior SCs.  
Prior tools address at most two of these SCs~\cite{chiou2023bagel,gdsa11yaudit,springer2025coverage}, and none evaluates focus obscuration (SC~2.4.11, SC~2.4.12) or focus appearance (SC~2.4.13).

WebAIM's WAVE~\cite{wave} is one of the most widely used accessibility evaluation tools. It uses rule-based checks on a page's HTML structure to annotate errors and ARIA landmarks. Despite its wide adoption, WAVE automates checks for only 13 of WCAG's 86 success criteria~\cite{wave,gena11y_he_fse2025,wcag22_guideline24}; the remainder require manual confirmation.

Axe-core~\cite{axecore} by Deque Systems targets zero false positives by design, ensuring every reported violation is genuine. Every rule fires only when a failure can be confirmed from the static DOM. This design means it systematically under-reports, leaving many violations undetected.

Two widely-used tools wrap axe-core for different deployment contexts. Lighthouse~\cite{lighthouse}, built into Chrome DevTools, produces a 0--100 accessibility score and is widely used in CI (continuous integration) pipelines. Pa11y~\cite{pa11y} provides a command-line interface for integration into automated test suites. Both inherit axe-core's rule set and share the same limitation: coverage is bounded by what can be confirmed from the static page structure.

Among static tools, QualWeb~\cite{qualweb}, developed at the University of Lisbon, offers the broadest formally-specified coverage through the W3C Accessibility Conformance Testing~(ACT) framework. It targets SC~2.4.7 (Focus Visible) by inspecting CSS \texttt{:focus} declarations, the rules that control how an element looks when focused. JavaScript can override these styles at runtime. QualWeb cannot observe such overrides, so it misses focus indicators that are applied or removed dynamically.

Benchmarking studies confirm this coverage ceiling. One survey~\cite{springer2025coverage} found that the median tool detects only two unique violated success criteria. The UK Government Digital Service audit~\cite{gdsa11yaudit} confirms this finding.

The common limitation across these tools is reliance on static DOM analysis~\cite{springer2025coverage}. Evaluating focus-behavior criteria requires dynamic analysis, i.e., driving a real browser, simulating keyboard interaction, and observing the rendered page state at each step.

Some prior work has extended the evaluation by triggering interactions during testing. One study~\cite{fernandes2012} extended the WCAG~2.0 evaluation to Rich Internet Applications (RIAs, pages whose content updates dynamically via JavaScript) by firing JavaScript events and capturing DOM mutations (changes to the live page structure), exposing violations that appear only in post-interaction states. Another~\cite{durgam2023} detected accessibility anti-patterns called accessibility smells by recording interaction event sequences during screen reader sessions. Each addresses a narrow subset of criteria: structural markup in dynamic content or screen-reader-specific patterns. Beyond accessibility, other runtime approaches have targeted CSS layout failures through heuristic property localization~\cite{zerin2025heuristic} and retrieval-augmented repair~\cite{zerin2025repairing}, demonstrating the broader need for dynamic analysis of rendered web pages.

The most closely related work targets keyboard accessibility directly. An early approach~\cite{chiou2021} detected and localized keyboard accessibility failures by crawling pages and verifying that all interactive elements are reachable. BAGEL~\cite{chiou2023bagel} extended this with dynamic crawling and visual analysis of the rendered page layout to detect navigation barriers for SC~2.4.3 (Focus Order) and SC~2.4.7 (Focus Visible). BAGEL demonstrates that dynamic browser analysis can detect focus-behavior violations, but does not extend to measuring focus-indicator size or contrast, nor to detecting whether a focused element is visually obscured by overlapping content properties that require a different probing strategy entirely. A companion study~\cite{chiou2023dialog} targeted keyboard accessibility failures inside modal dialogs, including focus-trap violations (states where keyboard focus cannot leave a component) and incorrect focus restoration. All three are limited to one or two criteria and a restricted interaction scope: none evaluates focus-indicator size and contrast (SC~2.4.13) or focus obscuration by overlapping content (SC~2.4.11 and~2.4.12).

One study~\cite{llm_manual_to_auto} showed that LLMs can evaluate several WCAG criteria previously considered too complex for automation, by reasoning over rendered HTML. GenA11y~\cite{gena11y_he_fse2025} extends this to 37 criteria, sending rendered screenshots and HTML source to an LLM (GPT-4o)~\cite{openai_gpt4} and prompting it against each criterion. Across these criteria it achieved 94.5\% precision and 87.6\% recall. However, GenA11y scored 0\% recall on SC~2.4.3 and SC~2.4.7, the two focus-behavior criteria it attempted. This is not a model limitation; it is architectural. A screenshot captures a single visual state. Focus indicators and tab-order relationships emerge only during keyboard interaction and are invisible to any snapshot-based approach.

To our knowledge, NavA11y is the first tool to cover all five WCAG~2.4 focus-behavior success criteria. SC~2.4.11, SC~2.4.12, and SC~2.4.13 are unaddressed by any prior tool. 

\section{Threats to Validity}
\label{sec:threats}

The primary internal validity threat concerns the composition of $D_d$, which contains 16 author-constructed pages. These pages were constructed by the same team that developed the detection logic and may therefore be skewed toward violation patterns that the implementation already handles. The six GDS test pages~\cite{gdsa11yaudit} were not visible during tool development. NavA11y achieves 100\% precision and recall on all six independently, providing a partial check against this bias. A real production page may trigger unanticipated combinations of CSS overrides, dynamic overlays, and unexpected browser behaviors not present in the test set. Whether this produces false positives or missed detections cannot be determined from $D_d$ alone. The $D_r$ manual audit (Section~\ref{sec:results-dr}) addresses precision on production sites, but recall cannot be measured there without exhaustive ground-truth labelling of all violations on each site.

The 26 websites were taken from the Semrush top-website ranking~\cite{semrush2024}, a corpus of high-traffic commercial sites. Violation rates on other types of website, such as government portals, small-business pages, and academic institutions, may differ. Reusing the corpus from~\cite{gena11y_he_fse2025} supports reproducibility but inherits any selection bias toward large commercial sites present in that ranking. This threat is partially mitigated by the fact that detection thresholds are derived from WCAG criterion definitions rather than tuned to patterns observed in $D_r$.

NavA11y snapshots the set of focusable elements once at page load and does not monitor subsequent DOM mutations. On pages that employ asynchronous content loading or infinite scroll, elements inserted after the initial snapshot are never traversed. Scroll-triggered DOM changes between the pre-focus and post-focus snapshots can corrupt style comparisons, producing incorrect verdicts for all element-level SCs. This limitation affects both accuracy and scalability. The current evaluation partially mitigates this by excluding identified infinite-scroll artifacts from the labeling dataset (Section~\ref{sec:results-dr}). Future work will integrate MutationObserver-based~\cite{mdn_mutationobserver} monitoring to detect DOM changes between snapshots, adopt incremental analysis that re-snapshots affected elements after mutations settle, and implement synchronization barriers that pause traversal until pending network requests and layout reflows complete.

\section{Conclusion}
\label{sec:conclusion}

This paper has demonstrated that all five WCAG~2.4 focus-behavior success criteria can be detected automatically through dynamic analysis. The proposed approach analyzes accessibility violations through two phases. The first simulates a full Tab key traversal to capture navigation order for SC~2.4.3. The second directly inspects each focusable element's rendered focus state and obscuration for the remaining four SCs. Each detection threshold is derived from WCAG specification text or validated through empirical pilot studies, ensuring that each verdict is based on defined criteria rather than approximation.

On the 32 controlled test cases, our approach detects every violation and raises no false positives. On 26 production websites, SC~2.4.13 violations appeared on every site; most websites rely on browser-default \texttt{:focus} outlines (the built-in styles applied when no custom focus style is declared), which routinely fall below the required indicator thickness and contrast ratio. SC~2.4.12 violations affected 81\% of sites due to persistent overlays covering focused elements. A manual validation by two independent front-end engineers confirms 90\% precision on a six-site stratified sample, establishing that the reported violations represent true positive cases for keyboard users. In addition to the approach, this paper releases a labeled dataset~($D_d$) of 22 test pages covering all five focus-behavior SCs, including pass cases, boundary cases, and obscuration scenarios that are absent from prior public datasets. Future work should extend the analysis to dynamic DOM mutations (content injected by infinite scroll or modal dialogs) and multi-step interactive widgets, and evaluate coverage on authenticated flows.

\section*{\uppercase{Acknowledgments}}
This research is supported by a fellowship from the ICT Division, Government of Bangladesh; No-56.00.0000.000.052.33.0001.25-159, dated 16.04.2025.

\bibliographystyle{apalike}
\bibliography{references}

\end{document}